# Deep and optically resolved imaging through scattering media by space-reversed propagation

W. Glastre[*], O. Jacquin, O.Hugon, H. Guillet de Chatellus, and E. Lacot

*Centre National de la Recherche Scientifique / Université de Grenoble 1, Laboratoire Interdisciplinaire de Physique, UMR 5588,Grenoble, F- 3804*
*Corresponding author: wglastre@ujf-grenoble.fr*

We propose a novel technique of microscopy to overcome the effects of both scattering and limitation of the accessible depth due to the objective working distance. By combining Laser Optical Feedback Imaging (LOFI) with Acoustic Photon Taging (APT) and Synthetic Aperture (SA) refocusing we demonstrate an ultimate shot noise sensitivity at low power (required to preserve the tissues) and a high resolution beyond the microscope working distance. More precisely, with a laser power of 10mW, we obtain images with a micrometric resolution over ~8 transport mean free paths, corresponding to 1.3 times the microscope working distance. Various applications such as biomedical diagnosis, research and development of new drugs and therapies can benefit from our imaging setup.

Along the path of a laser beam through a turbid medium, ballistic photons are exponentially converted into multi-scattered photons. As a result, two types of imaging techniques across scattering media have emerged over the past decades: those based on multi-scattered photons and those using ballistic ones. Acousto-optic imaging [1,2] or diffuse tomography techniques [3,4] belong to the first family. The advantage of these methods is that multi-scattered photons reach larger depths (several transport mean free paths) but the drawback is the loss of the initial direction of incoming photons: as a result the spatial resolution is limited to the millimeter range which is too large for many applications. On the contrary setups based on the detection of ballistic photons like Optical Coherence Tomography (OCT) [5,6] or confocal microscopy [7] exhibit high optical resolutions (~µm). However the accessible depth is quickly limited by the small number of remaining ballistic photons and by the working distance of the objective.

In this paper, we present a new sensitive imaging technique belonging to the second family, which combines Laser Optical Feedback Imaging (LOFI, based on the ballistic photons detection) [8], Acoustic Photon Tagging (APT) [9] and Synthetic Aperture (SA) [10]. We demonstrate a shot noise limited [11] microscopy technique with an optical resolution enabling images beyond the working distance of the microsope objective, while the laser power is kept low for imaging biological media without damage.

LOFI is both an ultrasensitive laser autodyne interferometer and a confocal imaging technique combining the high accuracy of optical interferometry with the extreme sensitivity of class B lasers to optical feedback. In this autodyne method, the optical beating between a reference wave and the signal wave (the light back-reflected by the target) takes place inside the laser cavity. Because we choose the total round-trip frequency shift close to the relaxation frequency of the laser, we can benefit from an amplification of the optical beating inside the cavity by the laser gain. It has been previously shown that the detection is shot noise limited [11] even at low power (10 mW here). In the LOFI technique, the laser plays the role of both the emitter and the detector and as a result is self-aligned and thus easily transportable. An image can be obtained point by point by scanning the target with two galvanometric mirrors.

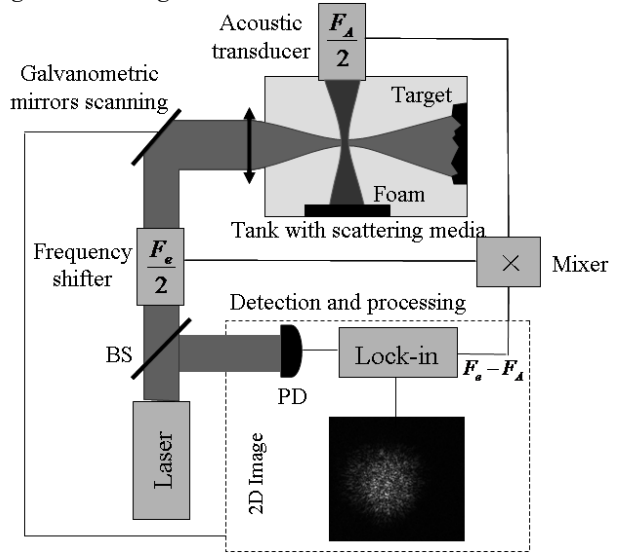

Fig. 1. Schematic diagram of Synthetic Aperture LOFI setup with acoustic tagging. PD, photodiode; BS, beam splitter; Fe and FA round-trip frequency shifts respectively induced by Acousto-Optic modulators and Acoustic transducer; × frequency mixer.

Our imaging system based on LOFI and SA is shown on Fig. 1. SA is a technique that can be used to make resolved images beyond the working distance of the objective, which is useful for making images deep in the medium. When a punctual target located at a distance L from the image plane of the objective is scanned, one obtains the point spread function [10]:

$$h_R(L, x, y) = \left| \exp\left(-\frac{x^2+y^2}{\left(\frac{\lambda L}{\pi r}\right)^2}\right) \exp\left(j2\pi \frac{x^2+y^2}{2L\lambda}\right) \right|^2 \quad (1)$$

This corresponds to a defocused image. However because both amplitude and phase of the reinjected electric field are accessible in LOFI, it is possible to numerically retropropagate the wavefront. This is the optical analogous to what was done with acoustic waves by Fink et al [12]. This technique is called SA operation and one finally get the following filtered signal [10]:

$$|h_{SA}(L, x, y)| = h_R(L, x, y) * h_R(-L, x, y) = \exp\left(-\frac{x^2+y^2}{r^2}\right) \quad (2)$$

It is thus possible to recover the same resolution r as if the target were initially in the image plane of the objective (see Fig. 2.)

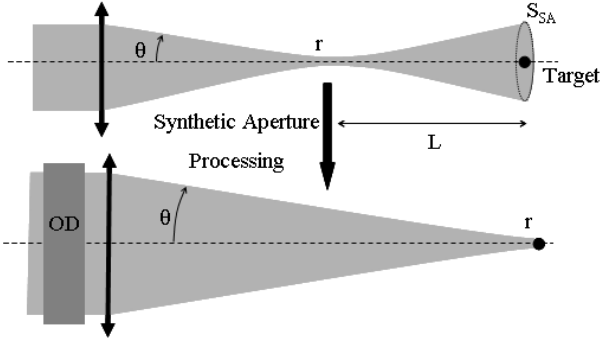

Fig. 2. Effect of Synthetic Aperture operation on the raw acquisition equivalent setup. L is the raw defocus, r the beam waist and θ the numerical aperture. OD = log(SSA/πr2) is the equivalent Optical Density and SSA the surface of the laser after a propagation L.

As can be seen from this figure, the major drawback of SA is the photometric balance: with the same noise level, compared to the classical (focalized) case, the signal power is decreased by a ratio $S_{SA} / \pi r^2$ (ratio between the surface of the laser beam in the plane of the target and in the image plane of the objective) [10]. Therefore the signal loss is proportional to $L^2$. LOFI is a confocal microscope used beyond the confocal zone [10] and can then quickly be limited by the background. More precisely, it was shown in [13] that the main limitations are parasitic reflections (see background on Fig. 3a and 3b) which are above the ultimate limit of the shot noise.

In order to eliminate the parasitic signal due to unwanted reflections we propose to use Acoustic Photon Tagging (APT) which eliminates the contribution of parasitic reflections [9] and offers real shot noise sensitivity giving access to a greater depth (L). Photons need to be tagged just before the target. Tagging is realized with an acoustic transducer which focuses an acoustic wave in the image plane of the objective (see Fig. 1). This acoustic wave produces a sinusoidal modulation of the pressure (amplitude ~MPa) at frequency $F_A/2 = 2.25$ MHz. As a consequence, the refractive index of the scattering medium and thus the optical path experienced by the laser are also modulated at the same frequency. The acoustic wave can be considered as a homogeneous phase modulator since its waist has a millimetric dimension which is the same order of magnitude than the field scanned by the laser. More precisely it induces the following phase modulation:

$$\Delta\Phi(t) \propto A\Delta n(t) = A\sin(2\pi F_A t / 2) \quad (3)$$

In this expression A is a coefficient depending on the acoustic wave (power, waist,..) and on the scattering medium. Eq. (1) then becomes:

$$h'_R(L, x, y, t) = h_R(L, x, y)\exp(j\Delta\Phi(t))$$
$$= h_R(L, x, y) \sum_{n=-\infty}^{+\infty} J_n(A)\exp(j2\pi n F_A t / 2) \quad (4)$$

In this expression $J_n$ is the $n^{th}$ order Bessel function. In order to detect only tagged photons, the signal can be demodulated (with the lock-in amplifier) at frequencies $F_e - nF_A/2$ instead of previous $F_e$ [10]. The elimination of parasitic reflection can then obtained at the price of signal loss related to the acoustic tagging. This efficiency in intensity (number of photons) is related to the square of the $n^{th}$ order ($|J_n(A)|^2$) of Bessel function (see Eq. (4)). To recover the phase one needs to consider the coherence between the lock-in amplifier demodulation and the optical beating inside the laser cavity. More precisely, a frequency mixer multiplies the references from RF drivers of the acousto-optics modulators (at frequency $F_e/2$) and of the acoustic transducer (at frequency $F_A/2$). After the mixer we get a reference at frequency $(F_e-F_A)/2$. This reference is used by the lock-in amplifier which demodulates the signal at a frequency corresponding to the double of this input reference, that is to say $(F_e-F_A)$. With this accurate reference, both amplitude and phase of photons tagged with an acoustic shift $F_A$ (n=2) are accessible. The tagging efficiency is then given by $|J_2(A)|^2$ and depends on the acoustic pressure.

With the complete setup of Fig. 1, we made an image of a target made of small silica beads of 30-40 μm diameter, located behind a hole of 1 mm diameter. This target was placed inside a rectangular glass tank filled with diluted milk acting as the scattering medium. This acquisition (fig. 3) was made with a lens of focal length f=75 mm, with a defocus L=5 cm and an initial waist r=20 μm. Without acoustic tagging one take $F_e = 4.4$ MHz and with acoustic tagging $F_A/2 = 2.25$ MHz and $F_e/2=4.45$ MHz. As a result, in both cases (with or without acoustic tagging), the total round-trip frequency shift is equal to 4.4 MHz which is near the relaxation frequency of the laser $F_R=3$ MHz. This frequency shift of 4.4 MHz is optimized in order to be both shot noise limited and far from saturation and non-linear effects [11].

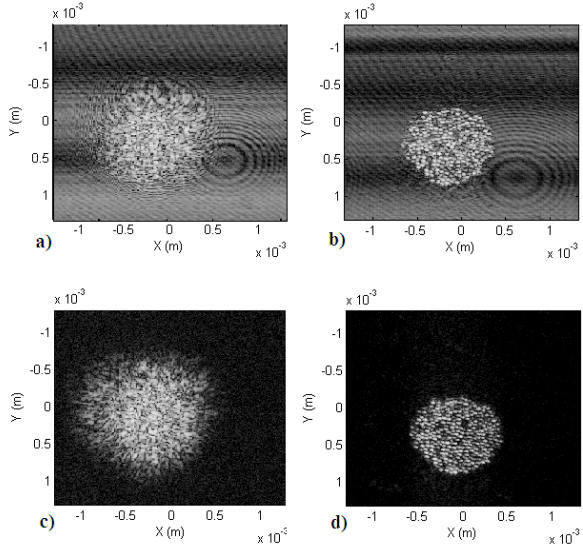

Fig. 3. Image examples (silica bead of 40 μm of diameter). a. and b. without APT; c. and d. with APT; a. and c. raw images; b. and d. images after SA processing. The total frequency shift is 4.4 MHz in both cases. The laser output power is 10 mW.

Comparing Fig. 3a-b with Fig. 3c-d clearly shows a double improvement: reduction of the background and recovery of an optical resolution. Fig. 3d shows that this setup enables to resolve beads after SA because a resolution r≈20 μm is recovered [10]: SA operation is now possible in combination to photon tagging. To give a more quantitative comparison without and with acoustic tagging, the concentration of milk is increased (resulting in the attenuation of the useful signal). Doing so, the useful signal is attenuated whereas the parasitic signal remains unchanged (because it arises before the scattering medium) [13]. Fig. 4 shows the experimental results: the so-called signal power is the sum of the squares (because the signal corresponds to a reinjected electric field) of all pixels values in the image.

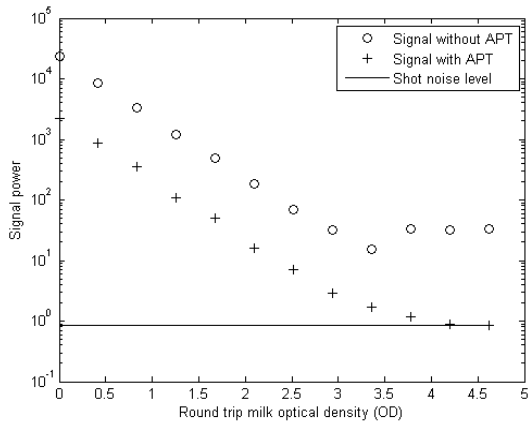

Fig. 4. Photometric performances with or without APT and comparison to shot noise. The total frequency shift is 4.4 MHz in both cases. Shot noise level is measured in the absence of reinjected photons.

From these results, one notes that without acoustic tagging, the image quality is limited by the background (above shot noise) and the target can be observed through a round-trip optical density of 3 (~ 7 transport mean free paths). On the contrary with APT we are limited by an acoustic tagging efficiency around 10% but the background is lowered by a factor ~40. That results in a capacity of making an image through an increased optical density of ~3.5 (~8 transport mean free paths) at the shot noise limit.

To conclude we have demonstrated the possibility to combine Acoustic Photon Tagging [9] with Synthetic Aperture [10]. Indeed possibility to get both resolved and shot noise limited (see Fig. 4) images through scattering media and beyond objective working distance has been proved. The acoustic tagging efficiency was limited in the present work to about 10% but could be improved by increasing the acoustic transducer power up to the maximum of $|J_2(A)|^2$ (around 20%). This would result in a higher sensitivity and an increased exploration depth. We intend to demonstrate this technique at the center of the therapeutic window (800 nm instead of 1064 nm) and with a both scattering and heterogeneous medium for a further extension to biological media.

This work was supported by a grant from the French Région Rhône-Alpes.